\documentclass[preprint,prm,preprintnumbers,amsmath,amssymb]{revtex4}
\usepackage{color}					
\usepackage{bm}					
\usepackage{graphicx}				
\usepackage{dcolumn}				

\begin{document}
\preprint{APS/123-QED}

\title{Bias-dependent diffusion of H$_2$O molecules on an Al(111) surface}

\author{Satoshi Hagiwara$^1$}
\email{hagiwara.satoshi@aist.go.jp}

\author{Chunping Hu$^2$}

\author{Satomichi Nishihara$^2$}

\author{Minoru Otani$^{1}$}
\email{minoru.otani@aist.go.jp}

\affiliation{
$^1$National Institute of Advanced Industrial Science and Technology (AIST), 1-1-1, Umezono, Tsukuba, Ibaraki
 305-8568, Japan \\
$^2$AdvanceSoft Corporation, 4-3, Kanda Suruga-dai, Chiyoda-ku, Tokyo 101-0062, Japan \\
} 

\date{\today}

\begin{abstract}
We investigate the process by which a water molecule diffuses on the surface of an Al(111) electrode under constant bias voltage by first-principles density functional theory.
To understand the diffusion path of the water on the Al(111), we calculated the minimum energy path (MEP) determined by the nudged elastic band method in combination with constant electron chemical potential (constant-$\mu_{\rm e}$) methods.  
The simulation shows that the MEP of the water molecule, its adsorption site, and the activation barrier strongly depend on the applied bias voltage. 
This strong dependence of the water diffusion process on the bias voltage is in good agreement with the result of a previous scanning tunneling microscopy (STM) experiment. 
The agreement between the theoretical and experimental results implies that accurate treatment of bias voltage plays a significant role in understanding the interaction between the electric field and the surface of the material. 
Comparative studies of the diffusion process with the constant total number of electrons (constant-$N_\mathrm{e}$) scheme show that the absence of strong interaction between the molecular dipole and the electric field leads to a different understanding of how water diffuses on a metal surface. 
The proposed constant-$\mu_{\rm e}$ scheme is a realistic tool for the simulation of reactions under bias voltage not only using STM but also at the electrochemical interface. 
\end{abstract}


\maketitle

 \section{Introduction} \label{s:intro}
Fundamental studies on metal/water interfaces \cite{henderson2002interaction,michaelides2003general, meng2004water,schnur2009properties} have attracted much attention because understanding the process of water reactions at the interface plays a central role in a wide variety of applications such as catalysis\cite{carrasco2012molecular} and fuel cells\cite{ogasawara2002structure}. 
A recent experiment to investigate the electrochemical interface reportedly showed that the molecular structure of water strongly depends on the electrode potential \cite{utsunomiya2014potential}. 
Furthermore, an experiment in which scanning tunneling microscopy (STM) was used found that the applied bias voltage affected the activation barrier of water diffusion on a Pt surface \cite{motobayashi2014adsorption}.  
To further understand the water diffusion processes and reactions, it is necessary to clarify the extent to which water adsorption and diffusion on the surface depends on the bias voltage. 

First-principles density functional theory (DFT) \cite{hohenberg1964inhomogeneous, kohn1965self} is a powerful tool for investigating the molecular adsorption and diffusion at the surface. 
However, previous theoretical studies on water diffusion processes were mainly carried out without the bias-voltage \cite{ranea2004water, ranea2012potential, karkare2015ab, rawal2015adsorption}. 
Generally, the electrode potential is related to the electron chemical potential ($\mu_\mathrm{e}$) of the electrode, and the difference in $\mu_{\rm e}$ between the two subjects defines the bias voltage. 
Thus, for a system to which a fixed bias voltage is applied, the value of $\mu_{\rm e}$ of the electrode (or sample) surface must be constant. 
Therefore, to control the bias voltage in the simulation, we need to control $\mu_\mathrm{e}$ \cite{lozovoi2001ab, tavernelli2002ab, schneider2014constant, benedikt2013modelling} beyond the constraint of a fixed number of electrons (constant-$N_{\rm e}$). 

The quantum mechanical theory formulated under a grand-canonical ensemble is indispensable for a simulation with the fixed $\mu_{\rm e}$ (constant-$\mu_\mathrm{e}$)\cite{lozovoi2001ab, tavernelli2002ab, schneider2014constant, benedikt2013modelling}. 
Two flexible simulation methods within the constant-$\mu_{\rm e}$ scheme have been proposed: The one is the fictitious charge particle (FCP) method developed by Bonnet et al. \cite{bonnet2012first}, and the other is the grand-canonical self-consistent field (GCSCF) method introduced by Sundararaman et al \cite{sundararaman2017grand}. 
These grand-canonical methods based on DFT can be applied not only to the electrochemical interface \cite{sundararaman2017grand, ikeshoji2017toward, sundararaman2018improving, haruyama2018analysis, weitzner2020towards} but also to a system subjected to a bias voltage such as in the STM experiment. 

In this study, we carried out a first-principles study of the adsorption and diffusion processes a water molecule undergoes on the surface of Al(111) by varying the bias voltage using the FCP and GCSCF methods. 
To understand the dependence of water diffusion on the bias voltage, we computed the minimum energy path using the nudged elastic band (NEB) method \cite{mills1995reversible, henkelman2000climbing, henkelman2000improved} under constant bias voltage. 
Specifically, in the case of aluminum, it is essential to understand the fundamental processes at the metal/water interface \cite{michaelides2004first,li2006symmetry,ranea2012potential}; thus, we used an aluminum electrode in our study. 

The paper is organized as follows:
Section ~\ref{s:method} provides a brief description of the basic ideas of the constant-$\mu_{\rm e}$ schemes, their computational procedures, and the computational details for the DFT and NEB calculations. 
In Sec.~\ref{s:results}, we discuss the results obtained by the NEB combined with the constant-$\mu_{\rm e}$ scheme. 
Finally, the conclusions of our study are presented in Sec.~\ref{s:summary}. 

\begin{figure}
\includegraphics[width=75mm]{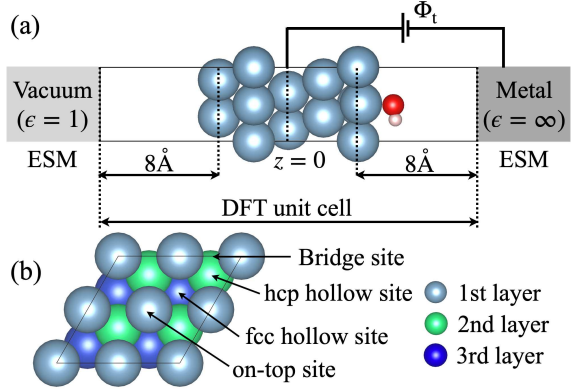}
\caption{\label{model}
(Color online). (a) Schematic view of the model under a bias voltage. The cyan, red, and pink spheres represent Al, O, and H atoms, respectively. The coordinate $z$=0 is defined as the center of the Al/H$_2$O system. 
Two ESM regions with dielectric constants $\epsilon$=1 and $\epsilon$=$\infty$, respectively, are attached to the left and right ends of the supercell to represent vacuum and the counter electrode (metal). 
(b) Adsorption sites on the Al(111) surface. The cyan, green, and blue spheres represent the first-, second-, and third-layer Al atoms, respectively, counting from the side of the counter electrode.  
}  
\end{figure}

\section{Methods and computational details} \label{s:method}

Here, we describe the computational methods and details. 
First, we provide an overview of the constant-$\mu_{\rm e}$ method under the boundary condition of the effective screening medium (ESM) technique \cite{otani2006first}. The next subsection presents a discussion of the computational procedures of the FCP and GCSCF methods. Finally, we provide the computational details of the proposed method.  

\subsection{Constant-$\mu_{\rm e}$ plus ESM method}

First, we briefly describe the basics of the constant-$\mu_{\mathrm{e}}$ scheme combined with the ESM method \cite{otani2006first}. 
The ESM technique, which was developed by Otani and Sugino, is a powerful tool for studying various material surfaces under repeated slab approximation. 
Figure~\ref{model}(a) shows a model of a constant-$\mu_\mathrm{e}$ calculation combined with the ESM used in this study. 
Two ESM regions with dielectric constants $\epsilon=1$ and $\epsilon=\infty$, respectively, are attached to the two ends of the supercell to represent the vacuum and the counter electrode (Metal). 
The electrostatic potential at the counter electrode was set to zero as the reference potential. Thus, the ESM enables us to always compare the energies measured from the same reference level.

To achieve the constant-$\mu_{\rm e}$ condition, we use the FCP \cite{bonnet2012first} and GCSCF methods \cite{sundararaman2017grand}.  
Both of these methods can be used compute the electronic structure and atomic geometry under the given target chemical potential $\mu_{\mathrm{t}}$. 
$\mu_{\rm t}$ is imposed by a potentiostat at a potential $\Phi_{\rm t}$, as shown in Fig.~\ref{model}(a), i.e. $\mu_{\rm t}=-e \Phi_{\rm t}$, where $e$ is the charge of an electron. 
We can define the grand-potential $\Omega$, instead of the total energy functional $E_{\mathrm{tot}}$, as follows:

\begin{equation}
\Omega=E_{\rm tot}-(N_{\rm e}-N^{0}_{\rm e})\mu_{\rm t}=E_{\rm tot}-\Delta N_{\rm e}\mu_{\rm t}, 
\end{equation}
where $\Delta N_{\rm e}$ is a fictitious charge particle, which is the difference between the total number of electrons in the system $N_{\rm e}$ and that in the neutral system $N^{0}_{\rm e}$. 
Then, we minimize the value of $\Omega$ for the atomic positions and $\Delta N_{\rm e}$, where $\Delta N_{\rm e}$ is not a constant but a dynamic variable during the entire minimization procedure. 
Because both the FCP and GCSCF methods converge to the same physical state, we expect these two methods to yield the same results under the same computational conditions.

\subsection{Computational procedure of constant-$\mu_{\rm e}$}

\begin{figure*}
\includegraphics[width=140mm]{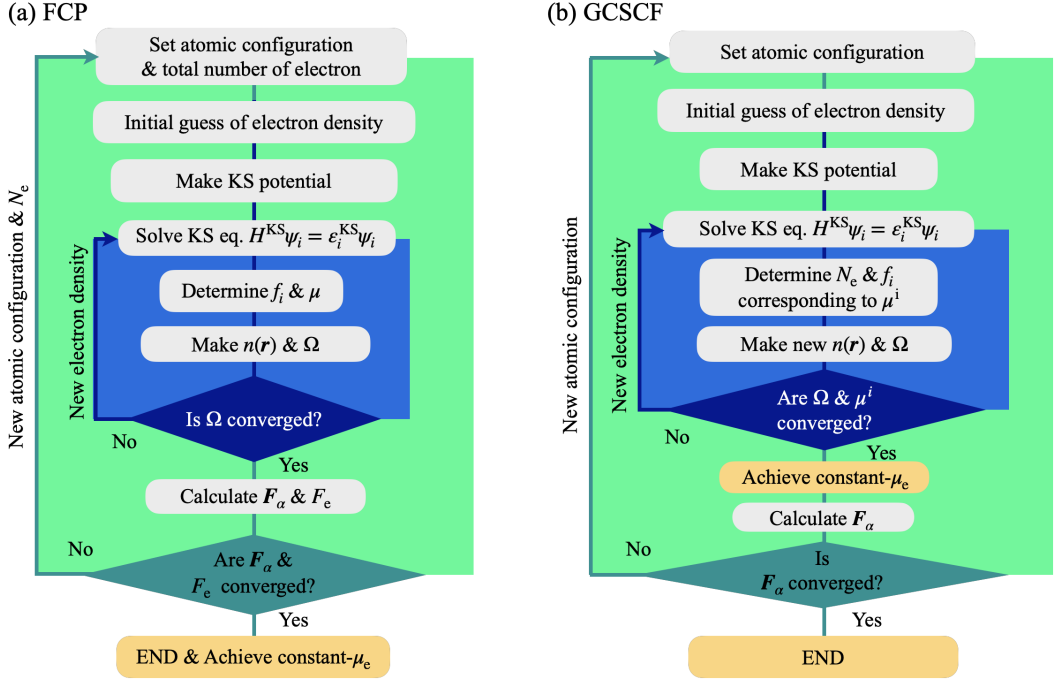}
\caption{\label{flow}
(Color online). 
Calculation flows for (a) FCP and (b) GCSCF methods. The green and blue shaded areas, respectively, indicate the calculation loops for the geometry optimization and self-consistent field. 
Here, $H^{\rm KS}$, $\varepsilon_i^{\rm KS}$, and $\psi_i$ denote the Kohn-Sham (KS) Hamiltonian, KS-eigenvalue, and KS-wavefunctions, respectively. 
The electron charge density is obtained by $n(\boldsymbol{r})=\sum_i f_i |\psi_i (\boldsymbol{r})|^2$, where $f_i$ is the occupation number for each state $i$. $\boldsymbol{F}_\alpha$ denotes the forces acting on each atom labeled by $\alpha$.
}
\end{figure*}

Here, we discuss the practical minimization procedure for $\Omega$ in the FCP and GCSCF methods. 
Figure~\ref{flow}(a) and (b), respectively, show the flow charts for the calculations with FCP and GCSCF, where we show a series of flows for the DFT calculation with geometry optimization. 
The SCF and geometry optimization loops show the blue and green shaded areas, respectively. 
In the following two subsections, we discuss the FCP and GCSCF methods using these flows. 

\subsubsection{FCP method}

The FCP employs a grand-canonical ensemble by the system connecting to the fictitious potentiostat, as shown in Fig.~\ref{model}(a), and minimizes $\Omega$ under the constraint of constant-$\mu_{\rm e}$ in the loop in which the geometry is optimized, which is shown as the green shaded area. 
In the SCF loop shown as the blue shaded area, the Kohn-Sham equation is solved with the fixed number of electrons.
Therefore, the system reaches constant-$\mu_{\mathrm{e}}$ via simultaneously optimizing not only the atomic positions but also the total number of electrons $N_{\rm e}$.
To optimize $N_{\rm e}$, we define a fictitious force for $N_{\rm e}$ as 
\begin{equation}
F_{\rm e}=-\frac{\partial \Omega}{\partial N_{\rm e}} = -\mu + \mu_{\rm t}, \label{omega}
\end{equation}
where, $\mu =\partial E_{\rm tot}/\partial N_{\rm e}$ implies the instantaneous $\mu_{\rm e}$, and yields the electrode potential as $\mu = -e \Phi$. 
To obtain $N_{\mathrm{e}}$ for $\mu_{\mathrm{t}}$, we minimize $F_{\mathrm{e}}$ using the quasi-Newton algorithm by using the Broyden-Fletcher-Goldfarb-Shanno minimization method \cite{broyden1970convergence, fletcher1970new, goldfarb1970family, shanno1970conditioning}. 
This method is commonly used for geometry optimization in conjunction with DFT calculations. 
In the geometry optimization with the quasi-Newton algorithm, the following equation updates all degrees of freedom for the atomic positions at the $k$-th iteration ($\boldsymbol{x}_{k}$), until all forces acting on the atoms become zero, as follows:
\begin{equation}
\boldsymbol{x}_{k+1}=\boldsymbol{x}_{k}+\boldsymbol{h}_{k}\boldsymbol{f}_{k}. 
\label{eq:newton}
\end{equation}
Here, $\boldsymbol{f}_{k}$ and $\boldsymbol{h}_{k}$ denote the forces acting on the atoms and Hessian, respectively. 
In conventional geometry optimization, the motion of $N$ atoms in three dimensions produces $3N$ degrees of freedom. 
In the FCP, both $F_{\mathrm{e}}$ and $N_{\mathrm{e}}$ are included in $\boldsymbol{f}_{k}$ and $\boldsymbol{x}_{k}$, respectively. 
Thus, we explore the solution of eq.~(\ref{eq:newton}) within the space of the $3N+1$-th dimension.  
However, we cannot directly treat $F_{\mathrm{e}}$ and $N_{\mathrm{e}}$ on equal footing with $\boldsymbol{f}_{k}$ and $\boldsymbol{x}_{k}$ because the units of $N_{\mathrm{e}}$ and $F_{\mathrm{e}}$ are different from the atomic positions and forces. 
To address this, we introduce the effective charge position ($N'_{\mathrm{e}}=\alpha N_{\mathrm{e}}$), where $\alpha$ is a scaling factor unit in bohr/$e$. 
$F_{\mathrm{e}}$ is also scaled by $\alpha$ as follows,
\begin{equation}
F'_{\rm e} = (-\mu + \mu_{\rm t})/\alpha.
\end{equation}
Here, the definition of $\alpha$ is $\alpha=L_{\rm max}/V_{\rm max} C_0$. 
$L_{\rm max}$ and $V_{\rm max}$ are the upper-bound of the change in the length and voltage at each step, respectively. 
$C_0$ is the capacitance determined by the formula of the parallel-plate capacitor: 
\begin{equation}
C_0 = \frac{1}{4\pi} \frac{S}{L}. 
\end{equation}
Here, $S$ denotes the surface area. 
Although the original definition of $L$ is the distance between the parallel plates of the capacitor, we approximately use the half-length of the unit cell in the z-direction for convenience. 
In the FCP optimization, we add $N'_{\rm e}$ and $F'_{\rm e}$ to $\boldsymbol{x}_{k}$ and $\boldsymbol{f}_{k}$, respectively. 

In eq.~(\ref{eq:newton}), $\boldsymbol{h}_{k}$ plays a role in determining the step width of not only the new atomic positions but also the new charge $N_{\mathrm{e}}$. 
For the Hessian component of $N_\mathrm{e}$, we use the first derivative of $\mu$ with respect to the excess charge ($- \partial \mu / \partial N_{\rm e}$). 
In the present implementation, we use the approximate inverse of the density of states (DOS) at $\mu$ for $-\partial \mu / \partial N_{\rm e}$ ($1/\rho(\mu)$). 
Generally, for a large DOS system near $\mu_{\mathrm{t}}$, $\mu$ gradually approaches $\mu_{\mathrm{t}}$ because of the small Hessian for $N_{\mathrm{e}}$ evaluated by $1/\rho(\mu)$. 
Thus, the convergence behavior of the FCP method depends on the DOS near $\mu_{\mathrm{t}}$. 

\subsubsection{GCSCF method}
Here, we briefly review the GCSCF method discussed in Ref.~\citenum{sundararaman2017grand}.
The GCSCF reaches constant-$\mu_{\rm e}$ during the SCF loop, is shown as the blue shaded area in Fig.~\ref{flow}. 
Because the formulation of GCSCF is simple, its calculation flow is essentially the same as that of the conventional DFT with geometry optimization. 
However, in the GCSCF, $N_{\mathrm{e}}$ is a variable at each SCF step. 
Generally, we can evaluate $N_{\rm e}$ by summing the occupied Kohn–Sham (KS) orbitals with the given $\mu_{\rm t}$. 
However, such a simple method for evaluating $N_{\rm e}$ violates the numerical stability of the SCF \cite{lozovoi2001ab}. 
Therefore, it is necessary to modify the numerical algorithms to determine the $i$-th transient Fermi energy ($\mu^i$) and update the electron density during the SCF loop. 

Now, we explain the algorithm for determining the $\mu^i$.
In the GCSCF, we gradually approach $\mu^i$ to $\mu_{\mathrm{t}}$ during the SCF as follows: 
First, we evaluate the Fermi energy $\varepsilon^i_{\mathrm{F}}$ at the $i$-th SCF step by the total number of electrons at the $i$-th SCF step, using an ordinal method. 
Second, $\mu^i$ is determined by simply mixing $\varepsilon^i_{\mathrm{F}}$ and $\mu_{\mathrm{t}}$ as follows: 
\begin{eqnarray}
\mu^i = \beta \mu_{\mathrm{t}} + (1- \beta) \varepsilon^i_{\mathrm{F}},
\end{eqnarray}
where $\beta$ is the mixing factor for $\mu_{\rm e}$ ($0<\beta<1$). 
Then, the total number of electrons corresponding to $\mu^i$ is determined, and we finally update the occupation number and the electron density using $\mu^i$ and the total number of electrons. 

Next, we discuss the charge mixing scheme using the direct inversion of the iterative subspace (DIIS) method \cite{pulay1982improved} within the GCSCF framework. 
At the $i$-th SCF step, we update the electron density by solving the KS eq. with an input electron density $n^i_{\rm in}$, and then the updated electron density is used as a $n^{i+1}_{\rm in}$. 
However, to obtain a more appropriate value of $n^{i+1}_{\rm in}$, the updated electron density is mixed with the electron density of the previous SCF steps. 
To accelerate the convergence of the SCF, we usually use the optimized electron density obtained by the solution of the DIIS method as $n^{i+1}_{\rm in}$. 
Usually, to stabilize the DIIS acceleration, the metric and Kerker preconditioning operators \cite{PhysRevB.23.3082} $\hat{M}$ and $\hat{K}$ are introduced, and the DIIS method updates the electron density without altering the term of $n(\boldsymbol{G}=\boldsymbol{0})$, where $\boldsymbol{G}$ is the reciprocal lattice vector. 
In contrast, the GCSCF requires the total number of electrons to be updated in the SCF loop. 
Therefore, to update $N_{\rm e}$, we introduce the dumping factor of $Q_{\rm K}$ ($>0$) to $\hat{K}$ and $\hat{M}$ as follows:
\begin{eqnarray}
\langle \boldsymbol{G} |\hat{K}|\boldsymbol{G} \rangle &=& \frac{|\boldsymbol{G}|^2+Q_{\rm K}^2}{|\boldsymbol{G}|^2+q_{\rm K}^2+Q_{\rm K}^2}, \label{kerker} \\
\langle \boldsymbol{G} |\hat{M}|\boldsymbol{G} \rangle &=& \frac{4\pi}{|\boldsymbol{G}|^2+Q_{\rm K}^2}. 
\end{eqnarray}
Here, $q_{\rm K}$ is the original damping factor of the Kerker preconditioning operator. 
When $Q_{\rm K}$ is set to zero, $\hat{K}$ and $\hat{M}$ revert to their original values. 
By introducing $Q_{\rm K}$, $\hat{K}$, and $\hat{M}$ at $\boldsymbol{G}=\boldsymbol{0}$ becomes a finite value, and we can always update the total number of electrons in the SCF loop. 
Once both $\Omega$ and $\mu^i$ have converged, the ordinal geometry optimization procedure provides the new atomic positions. 

\subsubsection{NEB method combined with constant-$\mu_{\mathrm{e}}$}

Here, we briefly discuss the NEB method in combination with the constant-$\mu_{\mathrm{e}}$ methods. 
The NEB method \cite{mills1995reversible, henkelman2000climbing, henkelman2000improved} describes the minimum energy path (MEP) for a chemical reaction by combining the images of the first and final states. 
The MEP is determined by solving Eq.~(\ref{eq:newton}) until the forces acting on each image become zero. 
Thus, in the conventional NEB method, we explore the solution of eq.~(\ref{eq:newton}) within the space of the $3N \times N_{\mathrm{im}}$ dimension, which corresponds to the motion of N atoms in three dimensions with $N_{\mathrm{im}}$ images of the MEP. 
In the NEB combined with the FCP, eq.~(\ref{eq:newton}) is solved within the $(3N+1) \times N_{\mathrm{im}}$ dimension because we extended the optimization space, as discussed in the previous section. 
In contrast, when used in combination with the GCSCF, the NEB does not require special modification of the optimization procedure for the conventional NEB framework because of its straightforward formulation. 
In this work, we employ the Broyden method \cite{broyden1965class} as a quasi-Newton algorithm for determining the MEPs.

\subsection{Computational details}
Here, we provide the computational details of this study. 
All calculations were performed using the \textsc{Quantum Espresso} package \cite{giannozzi2009quantum, giannozzi2017advanced}, which is a DFT code within the plane-wave basis sets and the ultrasoft-pseudopotential \cite{vanderbilt1990soft} framework. We implemented the FCP and GCSCF routines in combination with the ESM method in the package. 
We used the five-layered slab model in the p($2\times2$) supercell to represent the Al(111) surface with a single water molecule, which corresponds to a 0.25~monolayer coverage, as shown in Fig.~\ref{model}(a)]. 
The two ESM regions with $\epsilon=1$ and $\epsilon=\infty$ are located at a distance of $\sim 8~\mathrm{\AA}$ from the outermost Al layers.

For convenience, the experimental lattice constant of the face-centered cubic Al bulk ($4.05\times 4.05 \times4.05~\mathrm{\AA}^3$) \cite{popovic1992lattice} was used to construct the surface slab. 
The cut-off energies for the wavefunctions and charge density were 40~Ry and 320~Ry, respectively. 
The exchange-correlation functional was the spin unpolarized version of Perdew--Wang 91 within the generalized-gradient approximation \cite{perdew1992atoms}. 
$\boldsymbol{k}$-point sampling used a $5 \times 5 \times 1$ mesh in the surface Brillouin zone, and the number of electrons occupying the volume was determined by the Gaussian smearing method with a smearing width of 0.01~Ry. 
We carried out the structural optimization until $\boldsymbol{F}_\alpha < 5.0 \times 10^{-4}$~Ry/Bohr with the bottom of three Al layers fixed at bulk truncated positions. 
In the FCP calculation, the convergence thresholds for $\Omega$ and $F_{\rm e}$, respectively, are set to $1.0 \times 10^{-6}$~Ry and $1.0 \times 10^{-2}$~eV. 
In the GCSCF calculation, we decrease the threshold of the convergence criteria for $\Omega$ to $1.0 \times 10^{-8}$~Ry, and used the Thomas--Fermi charge-mixing scheme\cite{PhysRevB.64.121101}. 
The NEB \cite{henkelman2000climbing, henkelman2000improved} calculation under the constant-$\mu_{\rm e}$ condition, which enables the determination of the MEP between two stable endpoints, was carried out to determine the activation barriers and diffusion paths of H$_2$O molecules at the surface. 
The NEB calculation was conducted with ten discrete images for the MEP with a path threshold of $0.03~\mathrm{eV/\AA}$. 
The activation barriers and diffusion paths converged well for the number of images and path threshold. 

\section{Results and discussions} \label{s:results}

Here, we discuss the results of water adsorption and diffusion on the Al(111) surface as calculated using the NEB method with constant-$\mu_{\rm e}$.  

\subsection{Adsorption site of water-molecule on Al(111)}
First, we briefly discuss the adsorption energy and $\mu_{\rm e}$ at the neutral Al(111) surface with a H$_2$O molecule for adsorption sites, as shown in Fig~\ref{model}(b). 
The calculation shows that the on-top sites are the most stable for H$_2$O adsorption, and the adsorption energy we obtained for these sites is 0.23 eV. 
These results are consistent with the previous DFT result \cite{ranea2012potential}. 
For the MEP, we considered H$_2$O diffusion from one stable on-top site to the next on-top site via a bridge site \cite{neb_note}. 
To apply the bias voltage $U$, we measure $\mu_{\mathrm{t}}$ from $\mu_{\mathrm{e}}$ at the potential of zero charges (PZC), which was $-2.91$eV. 
In this study, $U$ is defined as $U = (\mu_{\mathrm{t}} - \mu_{\mathrm{PZC}})/e$ \cite{U_note}, and the three values of $U$ are examined: $U=$ $0$V, $-1$V, and $+1$V, respectively. 
Here, $\mu_{\mathrm{PZC}}$ is $\mu_{\mathrm{e}}$ at the PZC.

For comparison purposes, we also carried out the constant-$N_{\mathrm{e}}$ calculations by evaluating the excess charges ($\Delta N_{\rm e}$) induced by the bias voltage. The results we obtained for $\Delta N_{\mathrm{e}}$ under $U=-1$V and $U=+1$V are $+0.033e$ and $-0.036e$, respectively. 
The difference in the absolute values of $\Delta N_{\rm e}$ between $U=1$V and $-1$V implies that the response of surface electrons to the applied bias potential deviates from the linear response regime. 
In the constant-$N_{e}$ calculation, we imposed the obtained $\Delta N_{\rm e}$ during the diffusion processes. 

\subsection{Activation barrier of water-diffusion on Al(111)}

Table~\ref{tab_neb_E} presents the results of the activation barriers $E_{\mathrm{a}}$ obtained by the NEB with the constant-$\mu_{\mathrm{e}}$ and -$N_{\mathrm{e}}$ schemes. 
First, we briefly compare the $E_{\mathrm{a}}$ obtained by the FCP and GCSCF methods ($E^{\mathrm{FCP}}_{\mathrm{a}}$ and $E^{\mathrm{GCSCF}}_{\mathrm{a}}$). 
The results of $E^{\mathrm{FCP}}_{\mathrm{a}}$ are almost the same as those of $E^{\mathrm{GCSCF}}_{\mathrm{a}}$. 
Because the FCP and GCSCF methods converge to the same physical state, this agreement between the methods is reasonable. 
Hereafter, unless otherwise specified, we discuss the results of $E_{\rm a}$ using those derived with FCP as being representative of the constant-$\mu_{\rm e}$ scheme. 

The result of $E_{\mathrm{a}}$ for the constant-$N_{\mathrm{e}}$ with $\Delta N_{\mathrm{e}}=0.000e$ is very close to that of the previous climbing image NEB\cite{ranea2012potential}. 
In contrast, constant-$\mu_{\mathrm{e}}$ with $U=0.0$V produces a much lower $E_{\mathrm{a}}$ compared to $E_{\mathrm{a}}$ by the  constant-$N_{\mathrm{e}}$ with $\Delta N_{\mathrm{e}}=0.000e$. 
Under $U=-1.0$V, $E_{\mathrm{a}}$ increases to 0.161eV, which is also smaller than the counterpart value of 0.184 eV obtained with the constant-$N_{\mathrm{e}}$ scheme with $N_{\mathrm{e}}=+0.033e$. 
By switching the value of $U$ from $0.0$V to $+1.0$V, $E_{\mathrm{a}}$ decreases from 0.092eV to 0.024eV. 
$E_{\mathrm{a}}$ with $U=+1.0$V is still lower than that determined by using the constant-$N_{\mathrm{e}}$ method with $\Delta N_{\mathrm{e}}=-0.036e$. 
The above comparison between the constant-$\mu_{\mathrm{e}}$ and -$N_{\mathrm{e}}$ methods shows that although they yield similar trends for $E_{\mathrm{a}}$ either as a result of applying bias voltages or introducing excess charges, the results are quantitatively quite different. 
Compared to the previous experimental data of H$_2$O-diffusion \cite{motobayashi2014adsorption}, our results of $E_{\mathrm{a}}$ by the constant-$\mu_{\mathrm{e}}$ are in good agreement with the observation that the $E_{\mathrm{a}}$ of water diffusion changes significantly with respect to the bias voltage. 
Thus, the simulation with the constant-$\mu_{\mathrm{e}}$ method plays an important role in reproducing the experimental conditions for a constant bias-potential. 

\begin{table}
\caption{\label{tab_neb_E}
The activation energies $E_\mathrm{a}$ (in eV) of 
H$_2$O diffusion on the Al(111) under bias voltages of $U= 
0.0$V, $-1.0$V, and $+1.0$V using the constant-$\mu_\mathrm{e}$ scheme. 
The results of $E_\mathrm{a}$ for the constant-$N_\mathrm{e}$ scheme 
with excess charges $\Delta N_{\mathrm{e}}$ are also listed. 
The superscripts of $E_{\mathrm{a}}$ by the constant-$\mu_{\mathrm{e}}$ scheme, respectively, denote the results obtained by the FCP and GCSCF methods.  
}
\begin{tabular}{rrrrr}
\hline
\multicolumn{3}{r}{constant-$\mu_\mathrm{e}$} & \multicolumn{2}{r}{constant-$N_\mathrm{e}$} \\
$U$  & $ E^{\rm FCP}_\mathrm{a}$  & $E^{\rm GCSCF}_\mathrm{a}$ & $\Delta N_{\rm e}$ & $ E_\mathrm{a}$ \\
\hline \hline
$0.0$ V    &  0.092 & 0.091 & $0.000 e$    & 0.143    \\
$-1.0$ V  &  0.161  &  0.161 & $+0.033 e$ & 0.184    \\
$+1.0$ V  &  0.024  & 0.028  & $-0.036 e$ & 0.095    \\
\hline
\end{tabular}
\end{table}

\begin{figure*}[htb]
\begin{center}
\includegraphics[width=140mm]{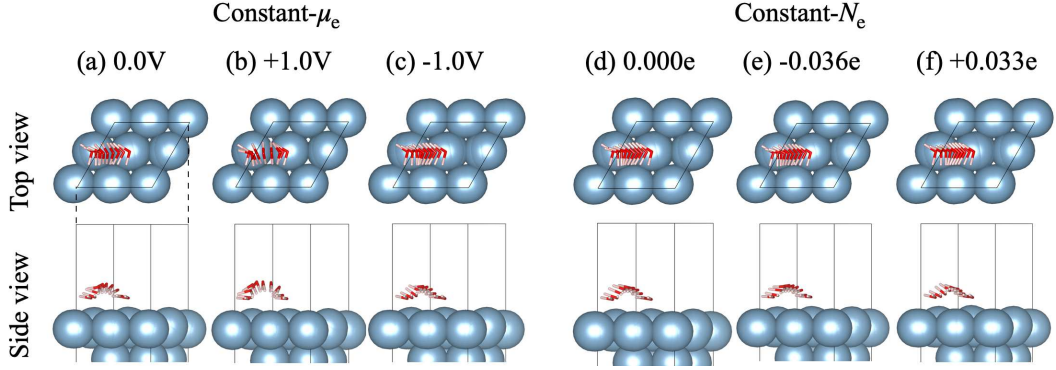}
\caption{\label{mep}
(Color online). Consecutive views of the MEP (ten images) for water diffusion on Al(111): results of constant-$\mu_{\rm e}$ scheme with bias-voltages of (a) 0.0V, (b) +1.0V, (c) -1.0V, and constant-$N_{\rm e}$ scheme with the excess charge of (d) $0.000e$, (e) $-0.036e$, and (f) $+0.033e$. 
The cyan, pink, and red spheres represent Al, H, and O atoms, respectively. 
The upper and lower panels show top and side views of the calculation cell, respectively.
}
\end{center}
\end{figure*} 

\subsection{MEPs for water-diffusion on Al(111)}

Here, we discuss the results of the water diffusion path obtained by the NEB with the constant-$\mu_{\rm e}$ and -$N_{\rm e}$ schemes. 
Figure~\ref{mep} (a)--(c), respectively, show consecutive images of the water diffusion along the MEPs at applied bias voltage of $U=0$V, $+1.0$V, and $-1.0$V, where we show the results obtained by the FCP as a representative example. 
Among the ten MEP images of H$_2$O, we regard the first and last images as identical, and the water molecules are bonded to the surface Al atom via the O atom. 
Along the diffusion path, the applied bias voltage drastically alters the dipole direction of the water molecule. 
We can evaluate the change in the direction of the H$_2$O dipole by obtaining the tilt angle between the dipole normal and the surface ($\theta$). 
For the first H$_2$O image, the values of $\theta$ under $U=0.0$V, $+1.0$V, and $-1.0$V are $76^\circ$, $85^\circ$, and $71^\circ$, respectively. 
This result indicates that the changes in the adsorption structure of H$_2$O resulting from the bias voltage are small for the first images of MEPs. 
However, the value of $\theta$ changes drastically in the intermediate images as a consequence of changes in the bias voltage. 
The dipole direction near the bridge site at $U=0.0$V and $+1.0$V becomes nearly perpendicular to the Al surface, and the H atoms orientate themselves downward. 
In contrast, the dipole directions for all images tend to be parallel to the surface at $U=-1.0$V. 

Figure~\ref{mep}(d)--(f) shows the results obtained for the MEP with the constant-$N_{\mathrm{e}}$ scheme. 
Here, the first and last images of H$_2$O in (d), (e), and (f) are the same as those in (a), (b), and (c), respectively. 
However, the changes in $\theta$ in the intermediate images are much smaller than those in the constant-$\mu_{\mathrm{e}}$. 
Thus, we interpret the difference in the results of $E_{\mathrm{a}} $ between the constant-$\mu_{\mathrm{e}}$ and -$N_{\mathrm{e}}$, as  presented in Table ~\ref{tab_neb_E}, as the difference in the MEPs. 
Therefore, this dependence of the H$_2$O geometry along the MEP on the applied bias voltage indicates the importance of the interaction between the water dipole and external electric fields. 

\begin{figure}[htb]
\begin{center}
\includegraphics[width=85mm]{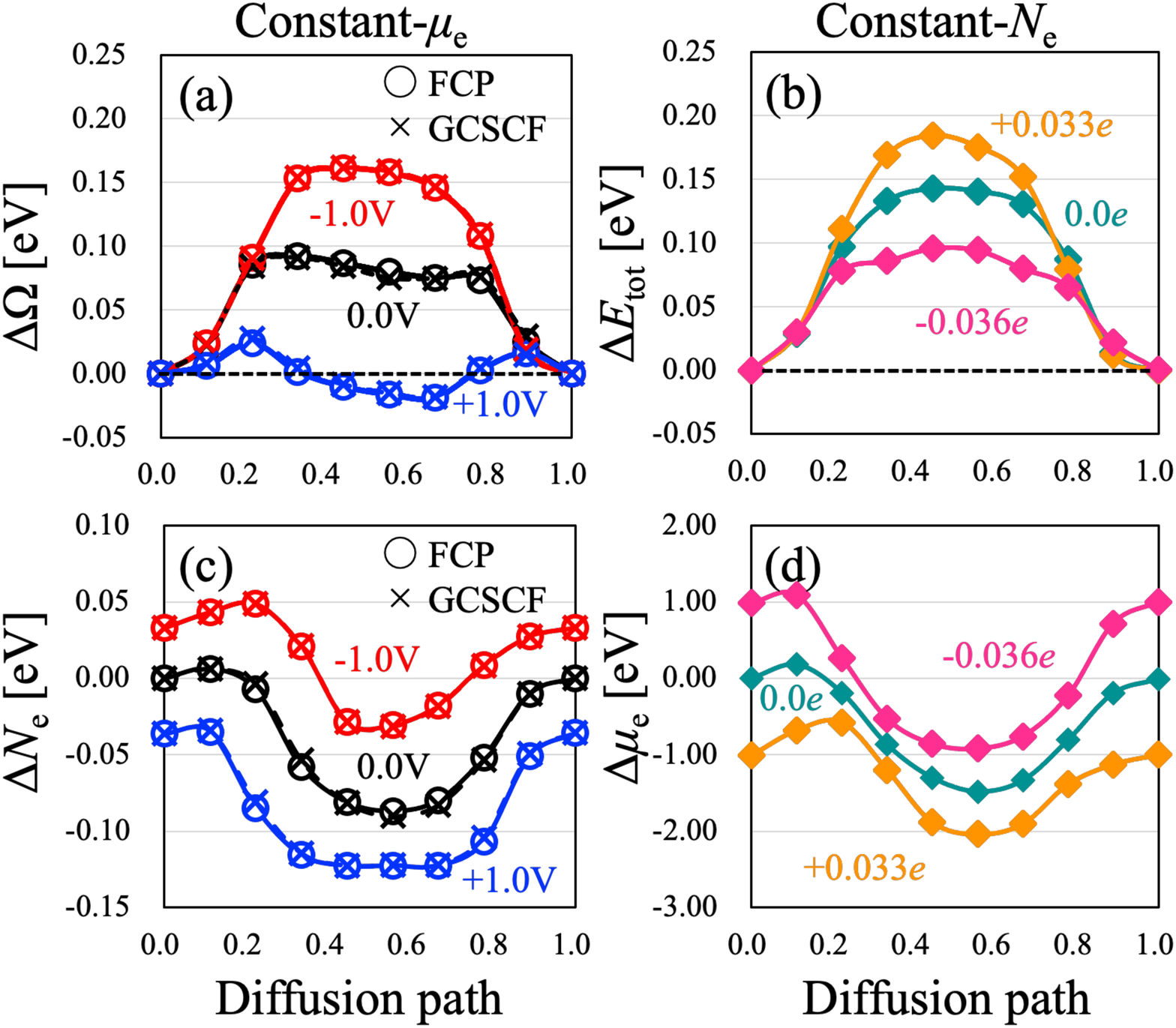}
\caption{\label{neb}
(Color online). Changes in the (a) grand-potential $\Delta \Omega$, (b) total energies $\Delta E_{\rm tot}$, (c) introduced excess electrons $\Delta N_{\rm e}$, and (d) chemical potential $\Delta \mu_{\rm e}$ as a function of the H$_2$O diffusion path. The results of (a) and (c) are obtained under $U=0.0$V, $-1.0$V, and $+1.0$V, and the results of (a) and (d) are obtained with $N_{\rm e}=0.000e$, $+0.033e$, and $-0.036e$. The open circles and cross symbols denote the results obtained by the FCP and GCSCF methods, respectively. The results of constant-$N_{\rm e}$ are represented by diamond symbols. The solid and dashed lines are intended to guide the eyes. }
\end{center}
\end{figure}

\subsection{Details of H$_2$O diffusion along the MEPs}
Figure~\ref{neb} shows the results of the analysis with NEB. 
Before discussing the details of the diffusion properties, we briefly discuss the results between the FCP and GCSCF methods shown in Fig.~\ref{neb}(a) and (c). 
Overall, the results of changes in the grand-potential $\Delta \Omega$ and excess charge $\Delta N_{e}$ along the MEPs obtained by the FCP (represented by open circles) and GCSCF (represented by cross symbols) are the same within the computational accuracy. 
These results indicate that we successfully implemented the constant-$\mu_{\rm e}$ methods. 
The main difference between the FCP and GCSCF methods is the computational procedure discussed in Sec.~\ref{s:method}. 
Because, as discussed above, we used the extended Hessian in the FCP method, the convergence behavior depends on the inverse of the DOS near $\mu_{\mathrm{t}}$.
In contrast, the GCSCF method directly optimizes $\mu_\mathrm{e}$ during a single SCF calculation. 
Because these constant-$\mu_\mathrm{e}$ methods employ different optimization procedures, we need to consider a different strategy to develop the FCP and GCSCF methods to more efficiently reach the constant-$\mu_{\mathrm{e}}$ condition.

Figure~\ref{neb}(a) presents the results of the $\Delta \Omega$, where the black, red, and blue circles, respectively, denote the values under $U=0.0$V, $-1.0$V, and $+1.0$V. 
In terms of the overall trend, the heights of $\Delta \Omega$ reach their respective maximum values in the intermediate images and decrease with increasing bias voltage. 
This behavior indicates that $E_{\mathrm{a}}$ of the diffusion of water on Al(111) depends on the external electric field, as listed in Table ~\ref{tab_neb_E}. 
For $U=+1.0$V, we found negative values of $\Delta \Omega$ at intermediate images of the diffusion path. 
This result indicates that the stable adsorption sites of the H$_2$O molecules on Al(111) change from the on-top sites to sites in the vicinity of the bridge sites. 
In contrast, the results of $\Delta E_{\mathrm{tot}}$ shown in Fig.~\ref{neb}(b) obtained by the constant-$N_{\rm e}$ scheme do not alter the sign of $\Delta E_{\mathrm{tot}}$ for any values of $N_{\mathrm{e}}$. 
Therefore, a comparison of the results of the constant-$\mu_{\mathrm{e}}$ and -$N_{\mathrm{e}}$ schemes would necessitate careful adjustment of the bias voltage to determine the stable adsorption site of H$_2$O on Al(111) in the STM experiments and at the electrochemical interface. 

Figure~\ref{neb}(c) shows the results of $\Delta N_{\rm e}$ along the diffusion pathway. In the first images, the values of $\Delta N_{\rm e}$ are the same as those used in the constant-$N_{\rm e}$ calculations. 
In the next few images, the value of $\Delta N_{\rm e}$ increases, and then it decreases in the intermediate images of the MEPs. 
This result is the consequence of introducing excess charges from an external potentiostat to Al(111) to maintain a constant bias voltage. 
The results obtained for $\Delta \mu_{\rm e}$ by using the constant-$N_{\mathrm{e}}$ scheme highly depend on the MEPs shown in Fig.~\ref{neb}(d). 
Here, we define $\Delta \mu_{\mathrm{e}}$ as the difference between the $\mu_{\mathrm{e}}$s in the MEPs and that in the first image of the MEPs of the PZC. 
Because the total number of electrons is fixed, this result originates from the charge transfer between the H$_2$O adsorbate and the Al electrode. This charge transfer alters the height of the dipole barrier for the substrate along with the MEPs. 
Because the change in the surface dipole barrier alters the work function that is directly related to the electrode potential, $\Delta \mu_{\rm e}$ highly depends on the H$_2$O diffusion path. 
Thus, these differences in the control mechanism of the surface charge between the constant-$\mu_{\rm e}$ and -$N_{\rm e}$ schemes provide the different MEPs for H$_2$O diffusion at Al(111), as shown in Fig.~\ref{mep}. 

\section{summary} \label{s:summary}

In summary, we demonstrated the realistic simulation of the bias-dependent diffusion of H$_2$O on the Al(111) surface using NEB calculations within the constant-$\mu_{\rm e}$ scheme. 
Our results showed that significant differences exist in the activation barrier energies and MEPs, and that this depends on whether the applied bias voltage is positive or negative. 
A comparison of the constant-$\mu_{\rm e}$ and -$N_{\rm e}$ schemes also showed that the conventional constant-$N_{\rm e}$ scheme does not provide a good description of molecular diffusion under a bias voltage owing to the absence of strong interaction between the molecular dipole and the electric field. 
In comparison, the FCP and GCSCF methods produced the same results within the computational accuracy. 
We expect the proposed scheme to find a wide variety of applications in the simulation of STM experiments and electrochemical reactions under constant electrode potentials. 

\begin{acknowledgments}
C.H. and M.O. thank Prof.~Osamu Sugino for valuable discussions. 
This work was supported by MEXT as the 
“Program for Promoting Research on the Supercomputer Fugaku” (Fugaku Battery \& Fuel Cell Project), Grant Number JPMXP1020200301.
The computations were performed using the supercomputers of Research Center for Computational Science (Okazaki, Japan), and 
Institute for Solid State Physics and Information Technology Center at the University of Tokyo. 
\end{acknowledgments}  


\begin{thebibliography}{45}
\bibitem[{neb()}]{neb_note}
\bibinfo{note}{The first and the last images of the NEB calculation were
  prepared by the most stable adsorption structure at a bias $U$. Then the
  intermediate images were initialized using the interpolation scheme both for
  the geometry and charge.}
\bibitem[{U_n()}]{U_note}
\bibinfo{note}{The sign of $U$ is set as in the STM: A positive sample bias
  drives the electron injected into the sample and increases the Fermi energy
  with respect to the $\mu_\mathrm{pzc}$.}
\end{thebibliography}

\end{document}